# Fault tolerant architectures for superconducting qubits

*David P. DiVincenzo*

*IBM Research Division, Thomas J. Watson Research Center, Yorktown Heights, NY 10598*

In this short review, I draw attention to new developments in the theory of fault tolerance in quantum computation that may give concrete direction to future work in the development of superconducting qubit systems. The basics of quantum error correction codes, which I will briefly review, have not significantly changed since their introduction fifteen years ago. But an interesting picture has emerged of an efficient use of these codes that may put fault tolerant operation within reach. It is now understood that two dimensional surface codes, close relatives of the original toric code of Kitaev, can be adapted as shown by Raussendorf and Harrington to effectively perform logical gate operations in a very simple planar architecture, with error thresholds for fault tolerant operation simulated to be 0.75%. This architecture uses topological ideas in its functioning, but it is not "topological quantum computation" -- there are no non-abelian anyons in sight. I offer some speculations on the crucial pieces of superconducting hardware that could be demonstrated in the next couple of years that would be clear stepping stones towards this surface-code architecture.

## Introduction

When I agreed to act as raconteur at this Symposium, the organizers created a title for my expected contribution, which the present article bears. I decided to accept the challenge of speaking and writing to such a title, even though it certainly implies a competency that I do not possess. There is a real specialty of "digital computer architect", and I believe that any actual practitioner in this specialty would find the thoughts about architecture that I will offer here unrecognizable as such. I offer them anyway because, well, we have to get started somewhere. I will argue that the experimental developments in the art of superconducting qubits, as reported at this Symposium, foretell a point in the future, not so immeasurably distant any more, when such a start will have to be made, or will be made for us. I think that when we begin to "architect" quantum processors, it will be right that we act anew, without close reference to digital computer architecture as it is practiced today. A quantum computer is a very different beast from a digital computer. I also think that as this endeavor progresses to maturity, it will finally recognize techniques from the classical discipline as valuable and applicable. But we will have to learn a lot before we can come to this point.

Here is the plan of this survey. First, I will briefly review the basics of quantum error correction. While it was already long before the discovery of quantum codes [1] that concepts for quantum computing machines [2,3,4] were first formulated, these cannot be considered sensible speculations, since the implementation of reliable operation is an essential defining characteristic of any computing machine. I will then briefly review the basic elements of fault tolerance. The discovery that fault tolerance was possible [5,6,7] did not initially provide a clear route forward, as the constructions of these papers did not incorporate some important physical considerations; for example, they were incompatible with any simple physical layout. But I will discuss recent developments that show how effective fault tolerance could be feasible in a two-dimensional layout. I will close with some thoughts about what next steps these current results suggest on the experimental road to superconducting qubits.

Before proceeding, I should qualify the architectural pathway that I explore here: it will be that of the general-purpose quantum computer, a universal, programmable device capable of the function of a quantum Turing machine; a technical way to say this is that the machine should be able to solve, in polynomial time, any problem in the computational complexity class BQP ("bounded-error quantum polynomial") [8]. Such a "universal" approach would not be the right path for the architecture of various special purpose machines. These would include the quantum repeater (because it would have to be specially adapted to the quantum input and output channels), analog quantum simulation engines, or possibly quantum (Grover) search machines.

## Quantum error correcting codes and fault tolerance

When Shor's quantum factoring algorithm was announced in 1994, there was still no compelling reason to believe that quantum computation was a physically realistic model for a computing machine. The model was that of an ideal, closed quantum system subject to classical time-dependent external control, on which perfect projective quantum measurements were possible. Real quantum systems deviate from this model, particularly in the fact that they are, at least to some degree, coupled to a quantum environment. In general, this coupling causes the evolution of a real quantum computation to deviate from the ideal one. If there were no mechanism for restoring the quantum computation to its intended track in the face of such departures from the ideal model, quantum computation would not be achievable in the lab. In the face of this, workers immediately sought for error correction strategies, and Shor soon found [4] a paradigm for quantum error correction codes (QECC) which form the basis of the realistic models of quantum computation that we use today.

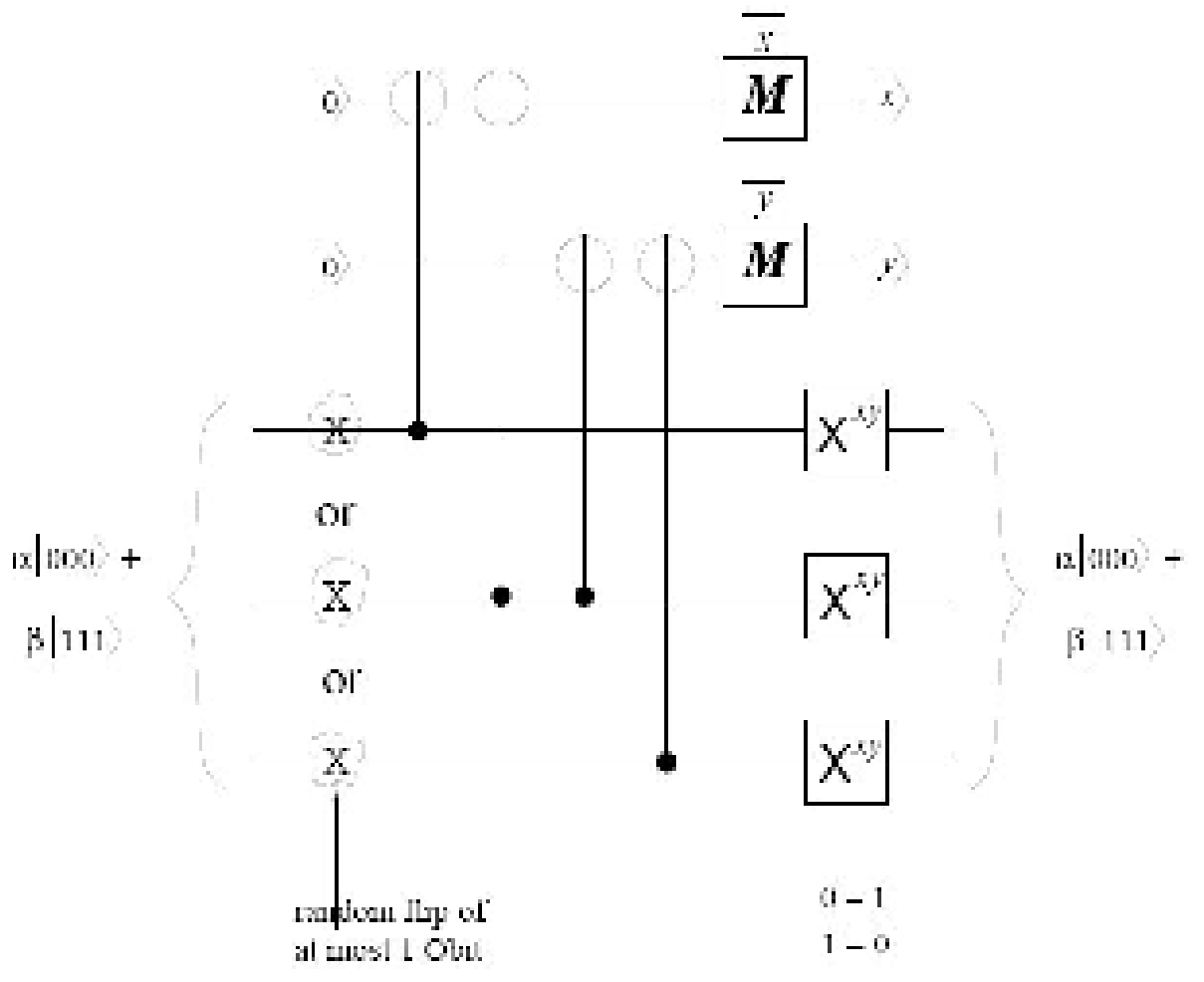

Fig. 1. Basic quantum circuit for the collection of syndrome information for the classical three qubit repetition code. The CNOTs accomplish two non-demolition measurements. The measurement outcomes are used as indicated to correct for the occurrence of X errors. Reprinted with permission from [9].

The first part of the QECC story, illustrated in Fig. 1, is that standard classical error correction codes are employable in a quantum setting. For, say, the three-bit repetition code, it is possible to learn about the occurrence of a single bit-flip error, and to correct for it, without disturbing the coherent superposition of the coded states. I don't want to dwell on this old story; I recommend the excellent chapter on quantum error correction in [9] for the reader who doesn't know all this already.

The next part of the story is to recognize that Fig. 1 is not enough, because the action of a quantum environment is more general than occasionally producing bit-flip errors (X errors) on the quantum bits. But it turns out that the most general action of a quantum environment, acting over a fixed interval of time, can be reduced to one of three errors, the Pauli-matrix operators, referred to as X, Y, and Z in this literature (see [10] for the details of measuring error rates). QECCs exist that correct for this larger but still discrete set of errors.

This useful but not immediately obvious mathematical fact is nicely explained in Sec. 5.3 of [9]. Actually, the description of quantum errors gets much more tricky if you consider correlations of the quantum environment from one time interval to others; QECCs are also effective in this "non-Markovian" setting, but the analysis of fault tolerance [11, 12] is more difficult in this case.

Shor [4] found a repetitive way of using the three-qubit code to satisfy these augmented quantum requirements, and after some time a huge taxonomy of such codes came into existence; a very specialized information theory literature has sprung up just on the mathematics of these codes. Among all these, one other code is worth mentioning, that due to Steane [13]. It is a seven-qubit QECC – the smallest such code that corrects a "general" (X, Y, Z) single-qubit error, and for which two-qubit operations can be done directly between coded qubits. "Directly" means that, for example, the CNOT between two logical blocks is accomplished by seven physical-level CNOTs between the corresponding members of the code block. For a long time, this style of "transversal" operation was the only known effective way of achieving error-protected gates between qubits coded using a QECC – but now there are new ways, as we will see later.

Figure 2 shows the error correction circuit, which has exactly the analogous function to the classical correction circuit of Fig. 1. It is close to being the simplest possible such circuit, but it must be admitted that "simple" is in the eye of the beholder. It definitely takes a lot more effort to correct one quantum error than it does to correct one classical error!

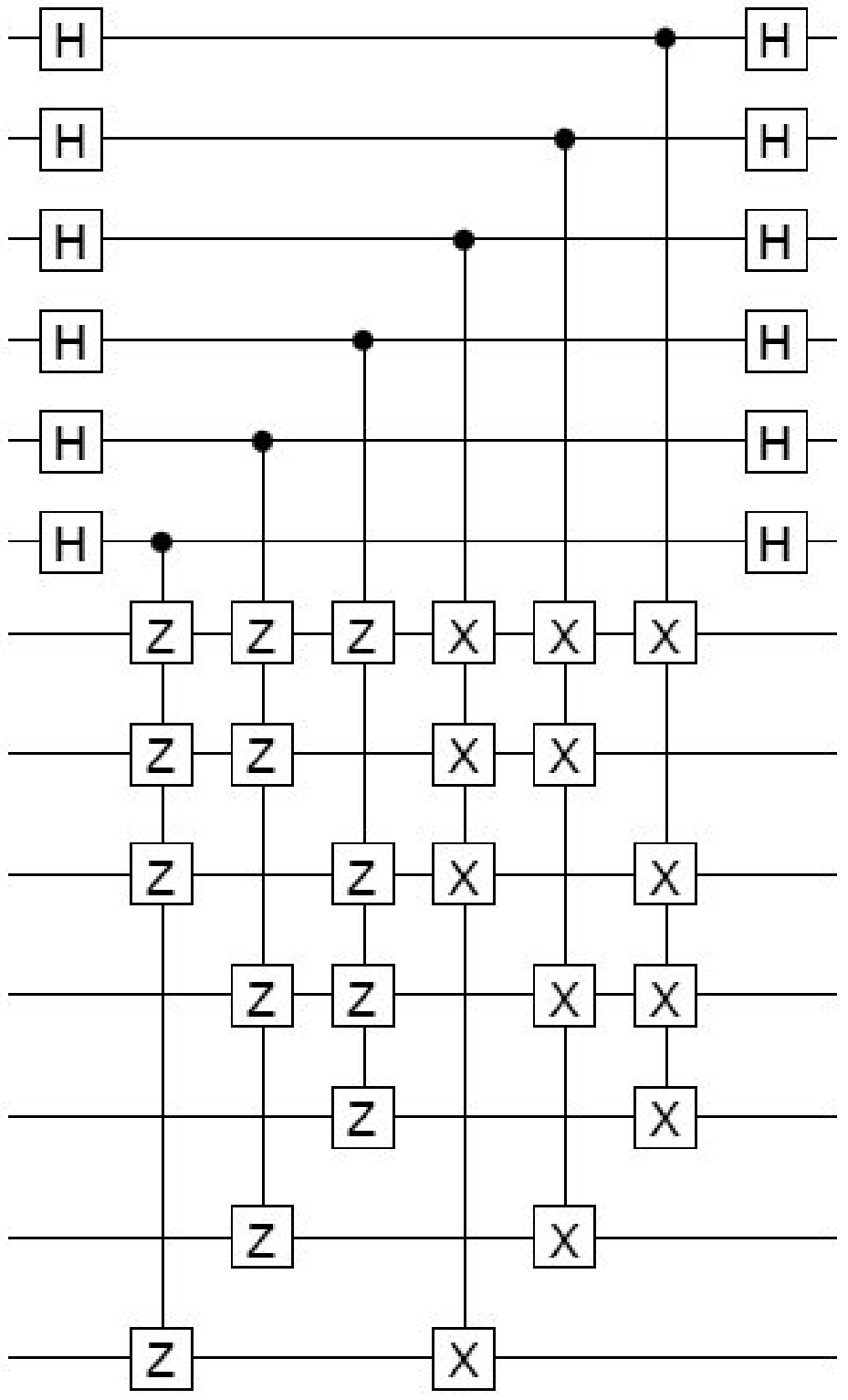

Fig. 2. Quantum circuit analogous to that of Fig. 1 for the correction of one general quantum error. The seven qubit code is that of Steane; the seven lower qubits contain the coded quantum bit. The upper six ancilla qubits collect the information necessary for six non-demolition measurements. Reprinted with permission from [9].

Unfortunately, the growth of complexity is not over: soon after such circuits were written down, it was realized that they are not *fault-tolerant.* For a code like Steane's that corrects a single error, the criterion for being fault tolerant would be this: assuming that *every element* of the error correction circuit could be faulty – not just the memory for the qubits before entering the circuit, but also the two-qubit gates, the Hadamard gates ("H" in Fig. 2), the ancilla preparation and measurement – the ideal state of the coded qubit should still be recoverable if a single error occurs in *any one* of the space-time positions in this circuit. It turns out that by suitable enlargement, and the addition of some extra "verifier" qubits, the error correction circuit for the Steane code, or for any QECC, can be made fault tolerant [7]. Effective suppression of noise has been seen in a limited sense in NMR experiments [14], but its demonstration in a modern, high-fidelity qubit would be a very significant milestone.

Of course, if a circuit is fault tolerant, it does not mean that error has been completely eliminated from its functioning. An error reduction is achieved, which we need to analyze. Consider a circuit for some quantum algorithm, which requires $T$ clock cycles and $N$ qubits. Without any error correction, we can estimate the overall probability of error as $TNp$, where $p$ is the probability of error in a single operation. It is easy to see that for a successful run of an algorithm of interest, requiring this overall probability to be <<1, absurdly low values of $p$ are needed (something like $10^{-15}$, for Shor-algorithm examples [15]). But with fault-tolerant encoding, the error probability can be reduced. For a code that corrects one error like Steane's, by doing error correction at every space-time location in the circuit, the effective error probability per operation becomes $Cp^2$, where $C$ is a constant which counts the number of distinct ways that two errors can occur in one round of error correction. $C$ is roughly $10^4$ for the Steane code.

This is still not satisfactory, since the error probability for the whole circuit, which now goes like $TNCp^2$, will still grow large for some algorithm size (that is, for sufficiently large values of $T$ and $N$). We must use codes that correct more errors. Fortunately, there is a variety of ways of obtaining codes that correct some greater number of errors $x$, for any value of $x$. For such a code, the probability of failure for an error-corrected qubit goes like

$$\textit{Failure probability of error-corrected qubit} = C[x]\, p^{x+1}.$$

Now, the strategy will be to choose a code, and therefore an $x$, with a foreknowledge of the size of the intended run of the algorithm $N$, and of the time $T[N]$ corresponding to that size. Thus, we consider $x$ to be a function of $N$ and write the failure probability of the whole algorithm:

$$\textit{Failure probability of coded algorithm} = T(N)NC[x(N)]p^{x(N)+1}.$$

We can hope that for any $N$, an $x$ can be chosen for which this probability can be made arbitrarily small, without too great a cost in coding overhead.

That this indeed works was proved in [5,6]. There is a real difficulty that needed to be dealt with, connected with the combinatorial factor $C[x]$. Unfortunately, for many families of known codes (including the "CSS" codes of which the Steane code is the smallest instance, and the "Bacon-Shor" codes of which Shor's original nine-qubit code was the smallest instance) this factor has a very fast increase [15] with $x$, something like $C[x] \simeq x^{cx}$ for some constant $c$. If nothing better were possible, the total error could not be made small by increasing $x$. But the proofs [5,6] found a workable solution, which involves creating code families by *code concatenation*. Concatenation is a code-construction method that iteratively replaces qubits in a code by their coded version. It is one concrete method to make a family of codes with growing $x$ and with a growth of $C[x]$ that is "only" exponential, $C[x] \simeq c^x$; in addition, the overhead of using these codes is acceptable, requiring in total $N poly(log(N))$ qubits. With this behavior of $C[x]$, there is a threshold value of the error probability $p_{th}=1/c$ below which the error probability can be made arbitrarily small, no matter what the value of $N$. When this is achieved by concatenation, $p_{th}$ is determined entirely by the properties of "small" codes like the Steane code above. But in this paper I will discuss another code family not created by concatenation, the surface codes, for which $C[x]$ is also well behaved [15]. This approach to fault tolerance is very successful, as I detail below.

# Fault tolerance and spatial layouts

Accompanying the proofs that quantum fault tolerance is possible [5-7], rough estimates were already given of threshold error rates $p_{th}$, that is, error rates below which failure probability of a long run of a full algorithm could be made very small. The numbers obtained in these studies (see also [16]), $p_{th}$ =$10^{-4}$-$10^{-5}$, remained canonical for a long time, even though they really were just connected with existence proofs and had no reason to be thought of as being a careful estimate of the best achievable threshold. One could argue that the reason that these numbers were generally accepted at face value is that they were so absurdly disconnected from what could be achieved in the lab that there was no point in carefully refining them. Nevertheless, such refinements were gradually made over the years, culminating in Knill's numerical study [17] which indicated that a threshold of $p_{th}$ =0.03 was reachable. It was understood in this and previous studies that achieving fault tolerance with any reasonable overhead would require error rates substantially lower (perhaps a factor of 5-10) than the threshold values. But 3%, or even lower, is an error rate we can now see within reach in the lab, so these analyses deserve further study. It might be mentioned that the question of how high this threshold number could go in principle has also received attention in the literature, with an absolute upper limit on the threshold of around 47% being reported [18] (but with no expectation that such a high number is actually achievable).

But, as also recognized from the beginning (especially in [7]), and going on all the way to Knill's work [17], these estimates are based on unrealistic views of how a quantum computer could be put together. One of the biggest issues was that it was assumed that at any moment in time, a two-qubit gate could be performed between any two qubits in the machine, without regard to their spatial locations. This patently unphysical assumption of nonlocality of operation quite rightly called into question the validity of all efforts that were made, including Knill's and many others', for estimating fault-tolerance numbers.

This deficiency, a primary roadblock to talking sensibly about machine architectures, was gradually addressed over time. Gottesman [19] was the first to indicate that concatenated error correcting codes were compatible with a real space layout, and that long-distance interactions would not be required (although it was assumed that the interactions were nonlocal enough to do all the steps for a small code as in Fig. 2). A finite threshold value could be proved in this setting, although the simple formula for estimating it indicated that smaller values might be required.

Gottesman's procedures were somewhat schematic, and several attempts were made to put them into more explicit practice. While [19] indicated that one- or two-dimensional local layouts were possible, it was found that schemes for putting qubits in a 1D or quasi-1D arrangements led to substantial added overhead, with reduced values of the fault-tolerance threshold ($p_t=10^{-7}$ in one study [20], very recently [21] improved to $p_{th}=10^{-5}$).

Two dimensions has turned out to be quite a different matter. We took up this question [22], with the object of having a completely explicit scheme for accomplishing all necessary operations on a two-dimensional lattice of qubits with couplings only possible between nearest neighbors. We found such a scheme employing the Steane code (Fig. 2) [23]; somewhat better results have now been seen for the Bacon-Shor code [24]. Figure 3 gives an idea of our constructions, which are found in full at [25]. The seven qubits of the coded data are indicated in red, with two logical qubits occupying two adjacent 6x8 regions of the lattice, with one spacer row in between. "Ancilla" qubits are in green, and additional qubits needed as "verifiers" are in blue. Indicated actions are to be performed in parallel: state preparation (P), various kinds of two-qubit gates and movement of qubits by swaps.

This does not look very efficient – many qubits are inactive, there's lots of empty space, and it takes 17 time steps before the data qubits even get involved in the circuit. Nevertheless, it works, and the threshold, around $10^{-5}$, is "only" an order of magnitude away from what could have been achieved by using the same basic coding, only relaxing the locality constraint.

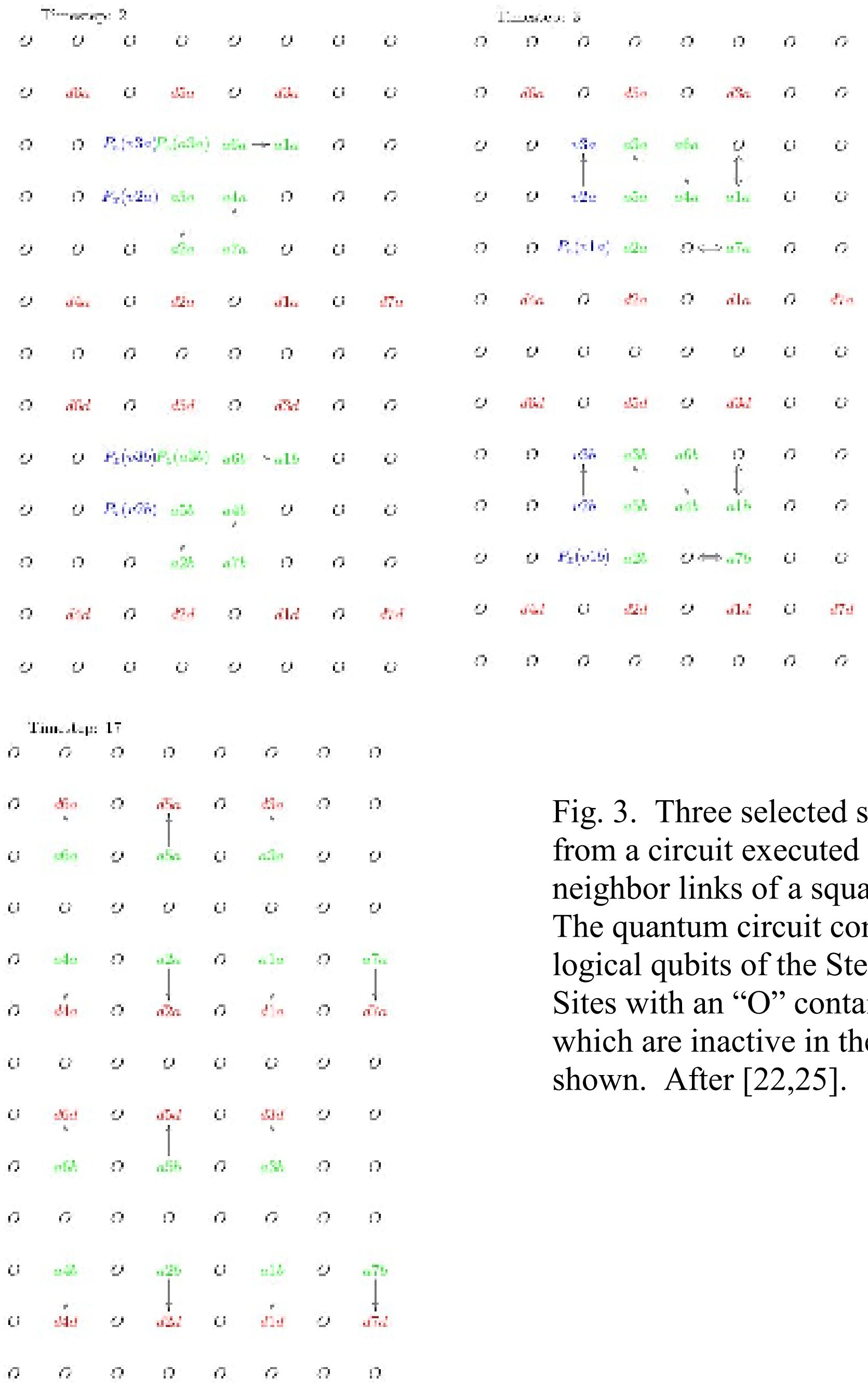

Fig. 3. Three selected snapshots from a circuit executed on near-neighbor links of a square lattice. The quantum circuit contains two logical qubits of the Steane code. Sites with an "O" contain qubits which are inactive in the timestep shown. After [22,25].

But from this proof of principle, we have now come a long way in the two-dimensional story. The story begins with the toric code of Kitaev [26], whose basic error-correction operations are shown in Fig. 4. The basic economy and simplicity of this lattice scheme compared with Fig. 3 is striking. All qubits are in use all the time, and the action is completely regular across the lattice.

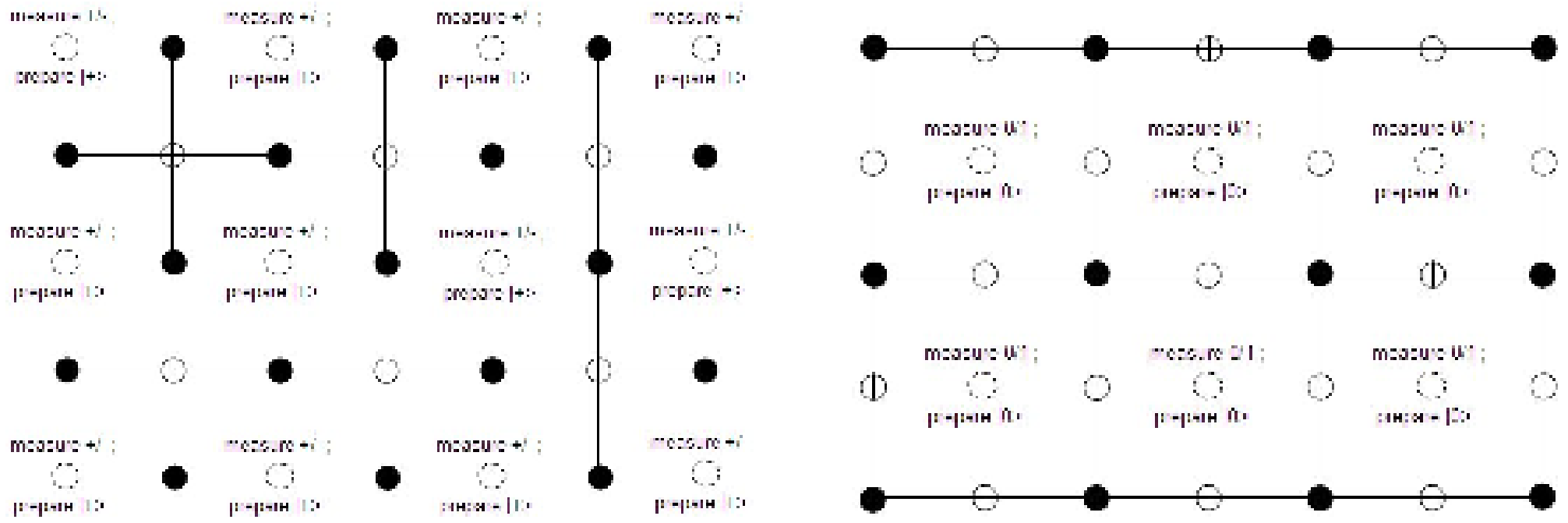

Fig. 4. Two successive timesteps of the circuit implementing the error correction for the toric or surface code. One quarter of the qubits are measured in every time slice, half of the qubits are never measured at all (except possibly at some later stage of operation). Simultaneous CNOTs shown in these figures all commute and so could be executed simultaneously.

But, some unsatisfactory things remain. As originally announced, this code was "toric", meaning that it involved periodic boundary conditions. But it was soon recognized [27] that various kinds of open boundary conditions were satisfactory; these are known collectively as the "surface codes". The next problem is that small versions of the code are evidently inferior to the 7-qubit Steane code: in fact, it requires a 13-site patch of square lattice for the surface code to correct one error. This is not a real problem, as it is only larger versions of these codes that are relevant for fault tolerance, but it perhaps prevented there from being an intense focus on the use of the surface code for some time. A serious study of the surface code family was undertaken by Dennis et al. [28], where a threshold near a percent was found for storing one qubit in this code. This paper established the important fact that the memory threshold corresponds to a phase transition in the statistical physics sense, one involving percolation of chains of undiagnosed or misdiagnosed errors from one boundary of the lattice to the other. This work did not yet imply such a high threshold for fault tolerant computation. Ref. [28] considered a scheme in which separate patches of surface code held coded qubits, with quantum gates performed by the lifting of one patch atop of another so that simultaneous coupling between every corresponding physical bit in the code ("transversal" operation) could be performed. Even with this 3D approach, the estimated threshold ($1.7\times10^{-4}$) for fault tolerant operation was not strikingly better than other estimates at the time.

But this is not the only way to use the surface code. It had already been recognized [29] that not just the boundary conditions, but also the topology of the patch of square lattice is relevant to the coded qubits that can be embedded in the lattice. In Fig. 5 we show, more schematically than in Fig. 4, a region of square lattice undergoing surface-code error correction. We imagine that the lattice is much bigger than is shown, but that it contains two holes. Holes are just places where we *refrain* from performing the error correction operation. In the example of Fig. 5, we show the smallest possible hole, which would involve just four of the CNOT gates coming in to one lattice point. But this pair of holes embodies a coded qubit. There is no definite answer to *where* this coded qubit is, but in a practical sense it is located in the immediate vicinity of the two holes: to do a Z operation on this logical bit, one does a chain of Z rotations of the qubits in a chain (any chain will do) surrounding one of the holes. To do a logical X, a chain of X rotations connecting the two holes suffices. So this means that a large number of hole-pairs, placed here and there in the lattice, will embody any number of logical qubits as desired. The quality of the coding is also easy to see: since undetectable errors in the logical qubits correspond to error chains as in Fig. 5, we see that if the holes are made big and far apart, the logical qubit is well protected.

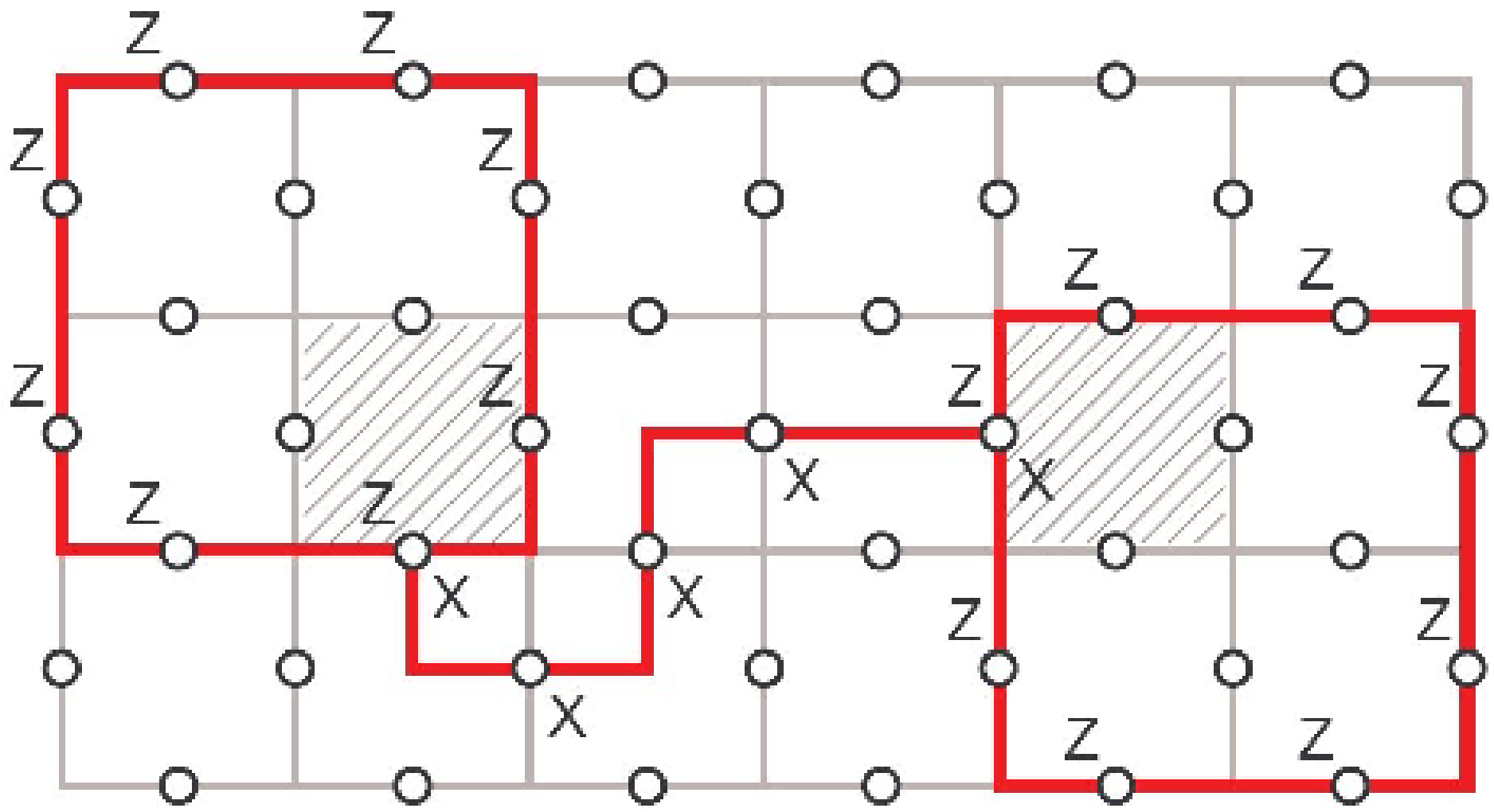

Fig. 5. Two holes in the surface code embody a logical qubit. The holes are obtained minimally from the surface code by leaving out four CNOTs coming to one qubit in two different places (see Fig. 4). The logical Z on this qubit is a loop of Z operations encircling one hole (either of the two shown will work), and a logical X is performed by doing a chain of X operations between the two holes. Reprinted with permission from [34].

The real break with previous work was the recognition of a new approach for doing logical gates by "code deformation" [32,33,30,28] (anticipated somewhat in "code teleportation" constructions, see Gottesman [16]), which does not rely on transversal operations. The scheme, as developed in [32,33,30], involves the braiding of holes [31]. A full and clear review of this scheme, with details explicitly worked out and explained, is given by Fowler et al. [34].

The important facts are these: First, holes, including large ones, can be moved via successive enlargement and contraction as shown in Fig. 6.

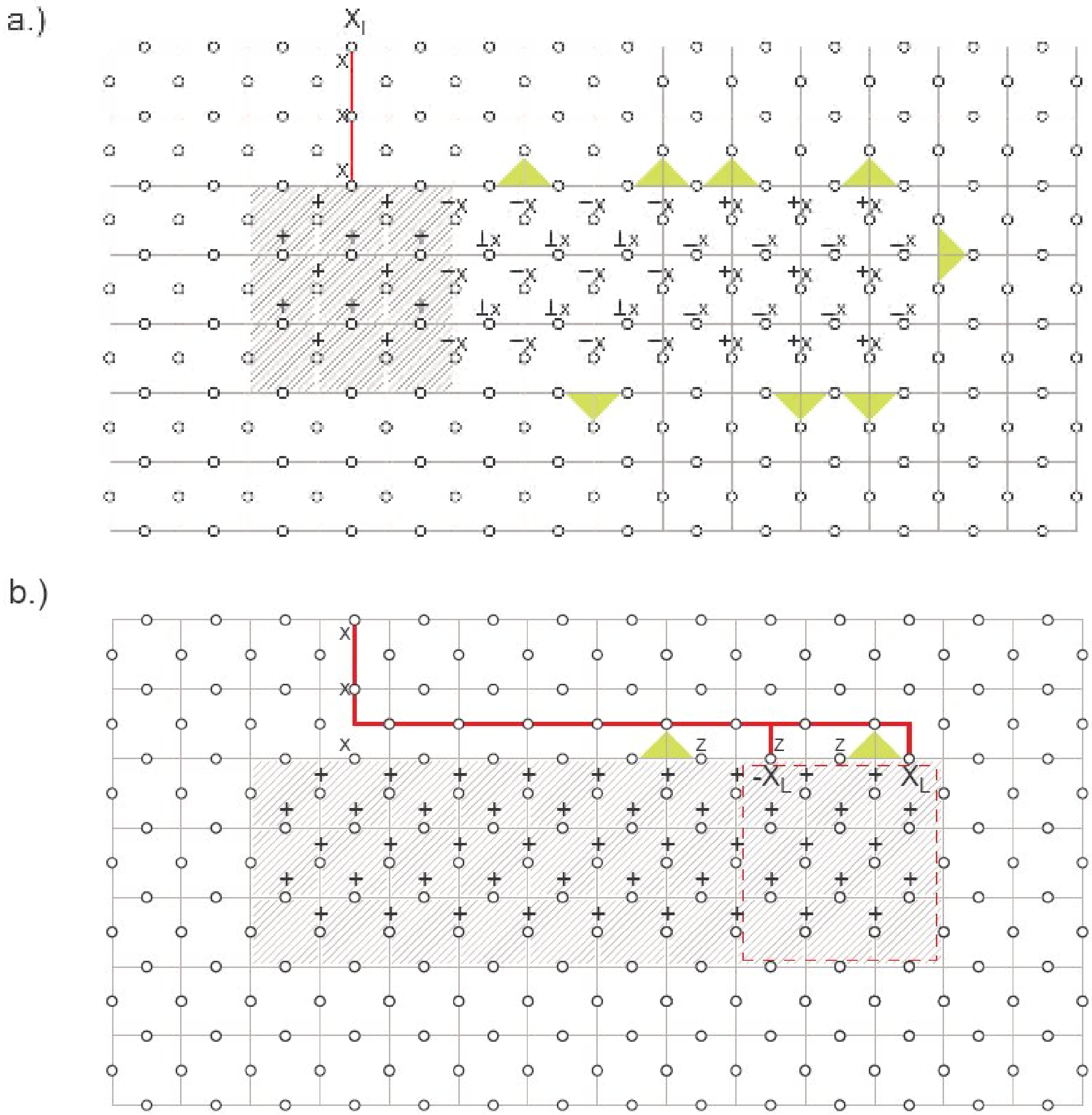

Fig. 6. Holes in the surface code can be large, which makes the logical qubits more protected from error. They can be moved from place to place by first extending it by measuring individual qubits in the patch ahead of the intended move, and then by resumption of error-correction measurements in the trailing region. Reprinted with permission from [34].

As shown, the logical operators (the Z and X chains) are dragged around when the hole is moved. Second, the ability to the correct errors for the logical qubit is not impaired, even when the hole is enlarged or contracted by a large amount. Third, when one hole of a pair is looped in between another pair, the dragging of the logical chains ($X_L$ in the Fig. 7) means that after this looping has been completed, this $X_L$ has been "copied" to the other qubit. (This figure uses the fact, which we will not describe in detail, that holes come in two dual forms.) This copying of Pauli operators (X from source to target, Z from target to source) is the defining property of a CNOT gate [16].

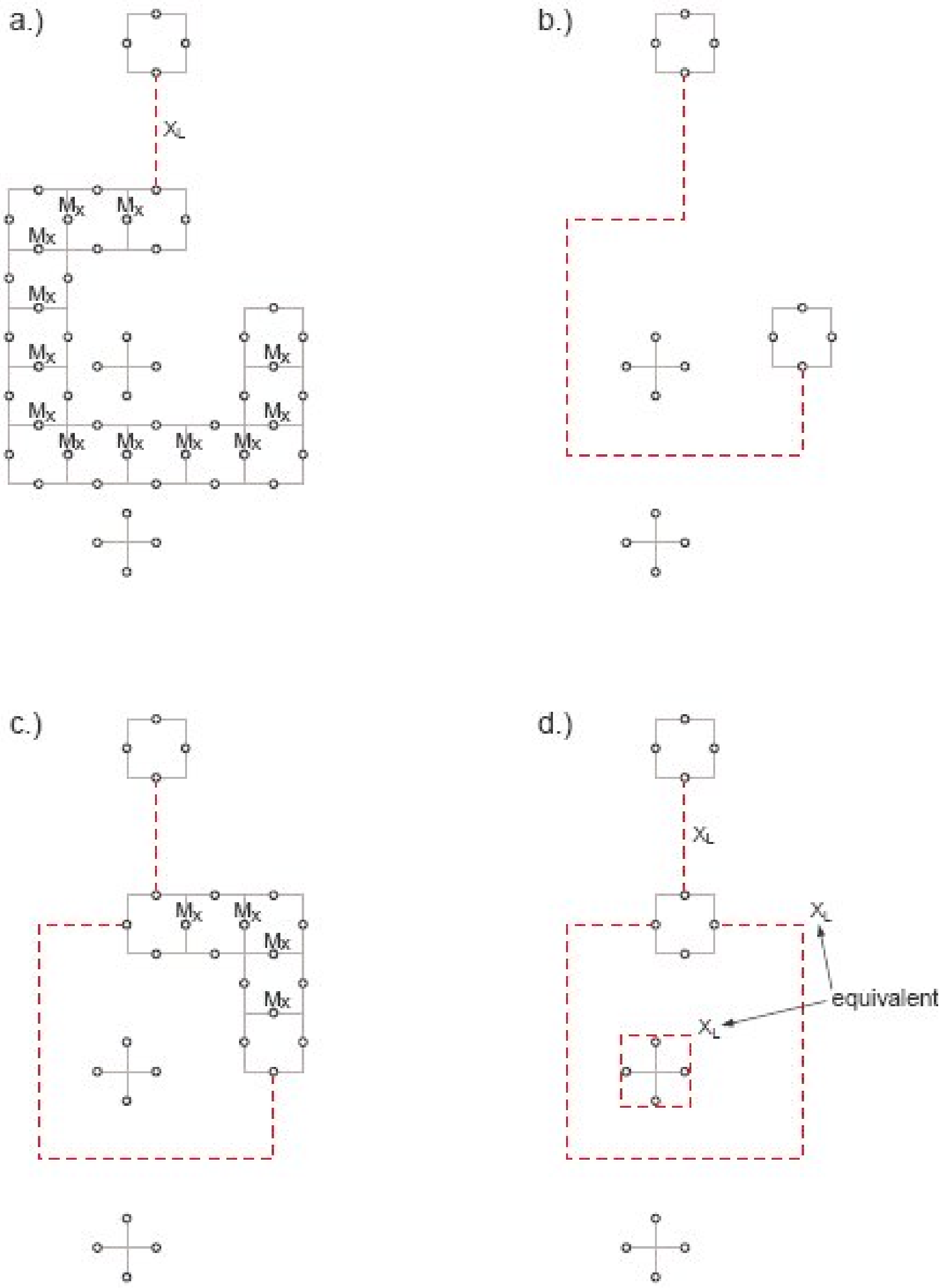

Fig. 7. CNOT is accomplished by braiding one hole from a pair around one hole of another pair. This is illustrated here for the qubits defined by the smallest possible holes. The braiding is accomplished quickly by the long-distance expansion and contraction of the holes along the braid path. Reprinted with permission from [34].

This method of doing logical gates via code-deformation braiding is very powerful, in that it permits error correction to go on continuously throughout the performance of the gate, so that the fault-tolerant computation threshold is the same as the threshold for reliable quantum memory. A set of other tricks are needed [33,34] to achieve a universal set of gates, but the determining factor for the threshold appears to be the CNOT trick. Simulations [33,35] indicate a noise threshold around $p_t$=0.008, close to the value reported by Dennis et al. for the memory threshold of the surface code.

These new approaches owe much to the developing ideas of topological quantum computation (TQC). Obviously, the central topological idea of gates by braiding has been the key development in the story told above. But, to the extent that the main focus of present work in TQC is on the search for states of matter that have topologically ordered ground states, the topological surface code development must be seen as a distinct species, a fruitful hybrid of the previously separate approaches to quantum fault tolerance. For intuitively understanding this scheme, it is permissible to use the language of Aharonov-Bohm effects that have been helpful in explaining the statistics of quasiparticles is topologically ordered phases. The measurement holes can be thought of as carrying various forms of electric or magnetic charge, and the process of carrying holes around one another induces state changes via an Aharonov-Bohm effect. Logical operators (e.g., the chain of X operations in Fig. 5) can be though of as operations that place new magnetic fluxes in the attached holes.

But, it should be noted that, at least in two dimensions, the active error correction used here, which identifies and removes errors, is superior to the Hamiltonian protection that is the hallmark of TQC [36]. So far, it is only in 4D that this passive approach is known to be effective [28]. The surface code scheme is definitely in the abelian category, and only in the most formal sense does it have anyonic particles, in the form of holes in which error-correction activity is absent. Our detailed understanding of its success as a technique for fault tolerant quantum computation rests firmly on traditional, combinatorial analyses of code performances. It is possible that other hybrids, owing more to the TQC approach, will come forth. For example, there are non-abelian extensions of the lattice-operator ideas, laid out by Kitaev [26] in his exposition of the toric code and explored further by Brennen et al. [37], which could lead to other successful fault tolerant schemes.

# A foreseeable architecture?

So, we see that qubits interacting locally on a square lattice are almost as effective as the most unrealistic interaction schemes for fault tolerant quantum operations. But, does this put the realization of quantum computers with superconducting qubits any closer to our reach? I think that forceful arguments can be made both on the positive and the negative sides of this question. Obviously, it is good to know that only short distance interactions are necessary, and that the CNOT alone, along with a small repertoire of local preparation and measurement operations, are all that is needed.

But many have thought that qubits in 1D is the only feasible way to go, for, it has been asked, for any of the plausible solid-state qubits, "Where will the wires go?" It is undeniable that the surface-code scheme assumes classical controls going to each and every qubit on the square lattice. The only conceivable answer is that these wires must come in from off the chip, in the third dimension. It was the opinion of my late IBM colleague Roger Koch that this solution should not put us off too much, that "three dimensional integration", which is presently very much in the air, will permit us to come to an architecture like this. It is also clear that no one is even remotely ready for such an extravagant piece of hardware, and it is fortunate that it is easy to conceive of controlling small patches of qubits in a square lattice without resorting to the third dimension.

I am very encouraged by developments in circuit QED, and I think that it is not absurd to take a very literal-minded view of the possibilities arising from coupling two qubits sitting near the ends of a transmission-line resonator. The qubits look like two vertices of a square lattice, and the resonator looks like a bond connecting them. The Deutsch-Jozsa implementation achieved recently by DiCarlo et al. [38] realizes this smallest piece of the square lattice, and has a two-qubit fidelity around 97% (3% error rate). 97% does not sound so far from 99.3%. Are we really this close to a scalable quantum computer?

No, naturally. But it is worthwhile to understand the other important ways in which we stand short of this goal. One big one is "quantum fanout", the number of other qubits with which a given qubit can participate in two-qubit gate interactions; of course, QF=1 in the two-qubit experiments of [38]. Getting to QF=4 to make Fig. 4 possible probably does not mean just adding qubits coupled to the same resonator (although see [39] for a possible approach). The literal-minded approach advocated above, which I think should be given a real chance, demands that each qubit must couple to multiple resonators – four of them, in the naïve approach. Mariantoni et al. [40] have done an extensive analysis of the first additional level of hardware complexity that is needed, in a variety of two-resonator, one-qubit simulations. (Ref. [41] advances related ideas.) They point out (see their illustration in Fig. 8) that such coupling can be either current- or voltage-based, is suitable for a variety of qubit types, and can involve a number of coupling geometries.

With some of these Mariantoni structures, higher interaction degrees are possible. One might think that QF=4 will require a qubit coupled to four independent resonators. It is possible to propose a device geometry for such a structure, although it is one that demands a rather complicated geometry for the qubit. Fortunately there is a much simpler structure in which QF=4 is be achieved, using the layout of Fig. 9 in which a single qubit is coupled in a "tee" arrangement to just two other resonators. With this structure, the surface-code functionality can be achieved with the skew-square tiling of Fig. 10. This figure shows the placement of two of the resonators and one of the qubits; completing these placements will give us sets of four qubits, around each small blue square tile, that can be made to function as a single qubit with suitable swappings, and new penalties in the fault-tolerance threshold. In explicit gate protocols we have worked out, at any given moment in time the surface-code qubit will be one of these four, and the other three will be functioning as ancilla. (Thus, there is no 4-qubit code involved in this construction.)

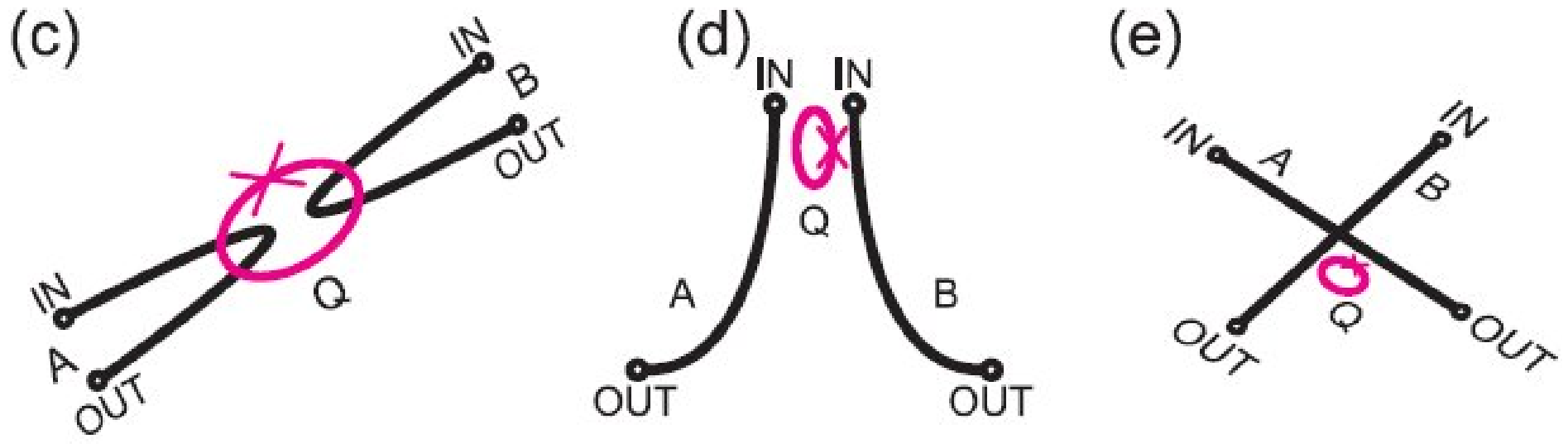

Fig. 8. Three distinct proposed schemes for coupling a qubit to two superconducting transmission line resonators. The coupling from qubit to resonator is strong, that from resonator to resonator is weak. Reprinted with permission from [40].

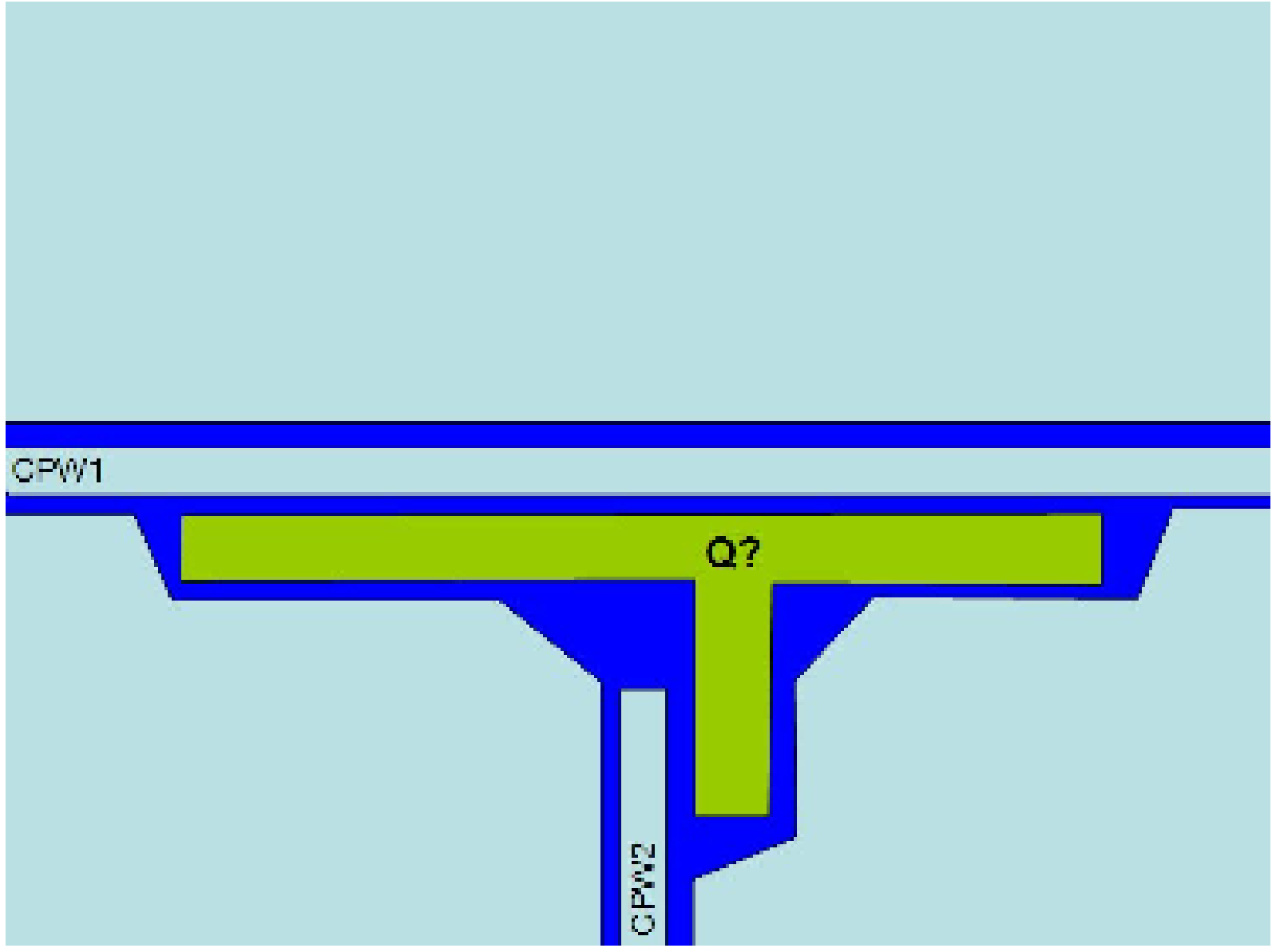

Fig. 9. A concept for a basic structure with one qubit coupled to two resonators. Dark blue is bare dielectric substrate, light blue is metal film constituting a “tee” structure of two coplanar waveguides (CPW), and green is a qubit of unspecified character (transmon, phase, flux, or other qubits could be made the embody the implied couplings). Many other elements (flux bias lines, measuring transmission lines) are not shown but could be made in a planar fashion.

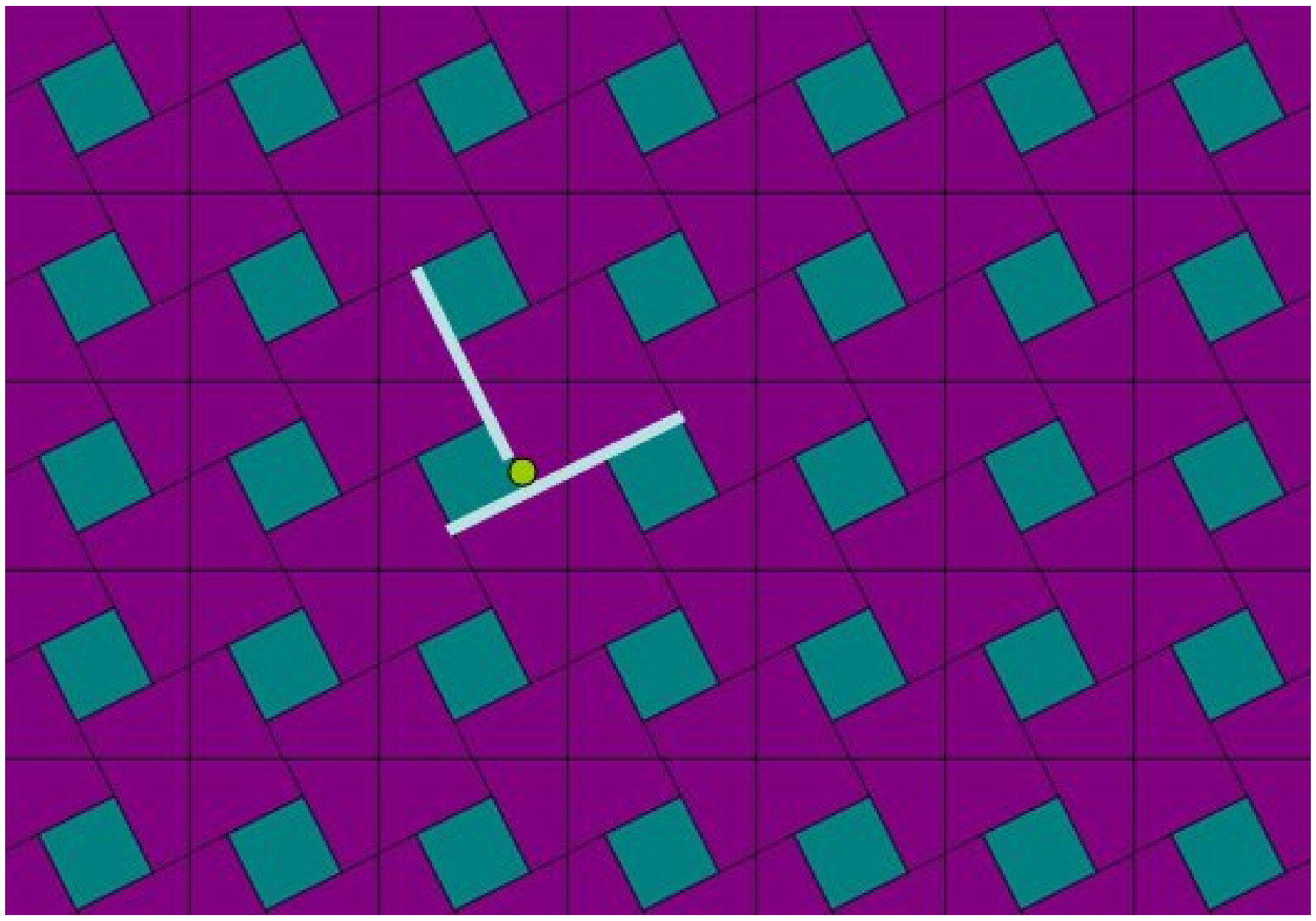

Fig. 10. A skew-square tiling on which the T structure of Fig. 9 could be repetitively placed to obtain a lattice that would be suitable for surface-code quantum computing.

Is this all that is needed? In the highest-level sense, yes, but with many, many things swept under the rug. My cartoons have not shown the many additional components that will be needed to control, prepare, and measure these qubits; the qubit of Fig. 9 may have to sprout more than the three legs shown to be functional, and other transmission lines will surely have to go somewhere. Another, rather simple, detail is that transmission-line resonators that are nearby should not have the same resonant frequencies, so that there are not unintended resonant couplings. A superlattice of different resonator lengths would suffice to solve this problem, as first anticipated in [2]. But more generally, there must be very rigorous suppression (at the 0.75% level, at least) of all unintended couplings and cross talk. All two-qubit operations should rigorously (at the same level, again) exclude entanglement with other qubits – a CNOT between qubits 1 and 2 should not involve nearby qubit 3. If this is not achievable directly, storage methods for qubits, in local two-level systems or by "parking" into other resonators [42], may be a feasible approach. Finally, we cannot make precise quantitative statements about the case when noise has a correlated or non-Markovian character, although we know that fault tolerance is still achievable in this more general case [11].

It is often emphasized that quantum computers are not analog computers, that because of error correction they function ultimately more like digital computers – noise, below a certain level, does not affect the successful course of computation. But this should not give comfort to the computer builder. Especially because of the small quantitative value of the threshold, the enterprise of building this machine will look, for a long time, like the construction of an extremely delicate, precise, complex, high-bandwidth analog device.

I should finally mention that, especially since this field is turning increasingly towards the manipulation of microwave photons, it is possible to consider architectural approaches that are being discussed for optical implementations of quantum computers. They indeed would have direct analogs in superconducting circuits. But unless the basic coherence of microwave photons in superconducting circuits is very much greater than the coherence of superconducting qubits, I don't think that the most popular of these approaches, the "linear optics" technique in which the Hong-Ou-Mandel effect is the central entangling phenomenon from which quantum gates are synthesized, will be superior to the qubit-oriented approach described above (but, see [43]). The overhead in the linear-optics approach is high, and the current noise-threshold estimate [44], even permitting spatial nonlocality, is below $10^{-4}$. It is possible that this outlook would be improved if the most recent developments in the topological-coding ideas are incorporated into analyses of linear-optics approaches.

To conclude, I have, as promised, plunged deeply into speculations about a future computer architecture based on current developments in superconducting qubits. Such speculations have come along regularly over the years, and it has been quite right for experimentalists, as well as for real, professional digital computer architects, to basically ignore them. Maybe the present speculations deserve the same, maybe not. But sooner or later, they will actually play a role in the further advancement of our field.

I am grateful to Austin Fowler for his careful explanation of the topological surface code scheme for quantum computation; it is his detailed and persuasive presentation that convinced me that it is the best approach that we have. Thanks to Barbara Terhal for extensive discussions, and for a careful reading of this manuscript. I thank the Institute for Theoretical Physics at the University of Amsterdam, for providing a hospitable location for the writing of this review; I am happy for support there from an ESF Exchange Grant under the INSTANS program. I am grateful for partial support from the DARPA QUEST program under contract number HR0011-09-C-0047.